**Why Super-Quantum, No-Signaling Correlations Cannot Exist**

**Pierre Uzan**

Fondation Santé des Etudiants de France

pierre.uzan@paris7.jussieu.fr

## Abstract

The idea that non-local correlations stronger than quantum correlations between two no-signaling systems could 'theoretically' exist is based on an incorrect statistical interpretation of the no-signaling condition. This article shows that any physically realizable no-signaling 'box' involving local incompatible observables indeed requires to be described in a non-commutative, quantum-like language of operators -which leads to the derivation of the Tsirelson bound and then contradicts this idea.

## Introduction

It is generally taken for granted in the literature that non-local correlations stronger than quantum correlations between two sub-systems that cannot exchange any signal are "theoretically possible" [1-3]. However, such no-signaling correlations that violate the Tsirelson bound have not yet been observed. Moreover, some indirect arguments have been formulated against the existence of these correlations, by mentioning the "multiplication of information" they would give rise to [7] or their rather implausible consequences regarding the cost of distributed computation [17]. It nevertheless seems that the absence of experimental evidence and the latter informational arguments are not enough to fully convince the supporters of the "theoretical possibility" of such correlations of their impossibility, and some of them explore very seriously their extraordinary consequences, like for example the possibility of non-local computation [1] [18]. The present article will explain why such super-quantum correlations cannot exist if the no-signaling condition is satisfied.

Like in the present debates about bipartite correlations, this question will be here discussed in terms of 'boxes'. A 'box', which is the central device of the Bell's game played by two parties [5], can be described by an arithmetic relation between couples of "inputs" (x,y), which can be regarded as the indexes of the two directions (right or left) each of the two parties (Alice and Bob) push her/his joystick, and "outputs" (a,b), which are the possible responses of the box for these actions. PR-boxes (labeled from the initials of Popescu and Rohrlich), compactly described by the following relation [1]:

$$P\ (a, b\ /\ x, y) = \tfrac{1}{2} \quad \text{if } a \oplus b = x.y \text{ is realized}$$
$$= 0 \quad \text{otherwise,}$$

where "⊕" is the addition modulo 2, are presented in the literature as paradigmatic theoretical models of no-signaling super-quantum correlations between Alice's and Bob's outputs.

Section I will briefly present the two assumptions on which the idea of existence of no-signaling super-quantum correlations relies and will emphasize the incorrectness of the statistical interpretation of the no-signaling condition, which thus calls into question this existence. Pushing further this investigation, it will be shown that the experimental sentences describing this Bell's game form a *partial* Boolean algebra reflecting the incompatibility of local actions (section II). As is well known from Birhkoff's and von Neumann's work on quantum logic and their recent developments [12-13], the latter structure is isomorphic to *a non-commutative language of projectors* onto the closed subspaces of an appropriate Hilbert space and not to a commutative language of functions on a phase space (section III). Consequently, the calculation of the Bell number *requires to be done within this non-commutative algebra of operators* -which leads to the derivation of the Tsirelson bound and then refutes the existence of no-signaling super-quantum correlations (section IV). The necessity of describing no-signaling boxes, including PR-boxes, in a quantum-like language of operators thus relies on the fact that such a description must clearly distinguish between two properties of the actions that can be performed by the players: on the one hand, the compatibility of Alice's and Bob's actions that can be realized conjointly, and on the other hand the incompatibility of Alice's alternative actions and of Bob's alternative actions.

**I. Could no-signaling super-quantum correlations be supposed to "theoretically" exist?**

Super-quantum, no-signaling bipartite correlations, which are supposed to "theoretically" exist [1][2], should violate the Tsirelson bound, which is the maximal amount of correlation allowed by quantum theory for no-signaling systems: $|R| > 2\sqrt{2}$, where the CHSH correlation factor R is defined by the following combination of correlators (here for the observables $A_i$ and $B_j$ respectively defined on the two sub-systems under consideration) [3]:

(1) $$R = C\ (A_0 , B_0) + C\ (A_0 , B_1) + C\ (A_1 , B_0) - C\ (A_1 , B_1).$$

They also should satisfy the no-signaling condition (NS), which intuitively asserts that no signal or no information can be transferred from one of the parties to the other. (NS) has been interpreted *statistically* in the literature [2-3] [8-9], by saying (Bell's game [5]) that the probability that Alice obtains a particular outcome "a" is independent of the choice of Bob's action "y", when he decides to push his joystick to the right or to the left -that is, this probability is independent of the value of y, and vice versa for Bob's result "b" and Alice's action "x". This statistical condition can be written in a condensed form as:

$(NS)_{stat}$ $$P(a / x, y) = P(a / x)$$
$$P(b / x, y) = P(b / y)$$

However, *this statistical interpretation* $(NS)_{stat}$ *of the non-signaling condition is incorrect*. The alleged equivalence between two assertions, namely (NS) and $(NS)_{stat}$, means that each time one of these assertions is satisfied the other is also satisfied, and each time one of them is falsified the other is also falsified. However, situations where (NS) and $(NS)_{stat}$ are not satisfied at the same time or refuted at the same time can easily be found [6]. For example, let us consider the case of a determinist box, for which $x_i$ determines the outcome $a_i$ for Alice and $y_i$ determines the outcome $b_i$ for Bob, and where Alice systematically informs Bob of her choice of action x. In this situation, $(NS)_{stat}$ is satisfied since $p(a_i/x_{i'},y) = p(a_i/x_{i'}) = \delta_{ii'}$ and $p(b_j/x,y_{j'}) = p(b_j/y_{j'}) = \delta_{jj'}$; but *(NS) is obviously not satisfied* since a signal is sent from Alice to Bob.

In fact, $(NS)_{stat}$ is only a necessary condition of (NS) since it is satisfied if (NS) is, but *it is not sufficient*, *which means that (NS) is stronger than* $(NS)_{stat}$. Consequently, even if $(NS)_{stat}$ allows the existence of no-signaling super-quantum correlations, *(NS), which is stronger, can forbid it* –which is indeed the case! Such a no-go result has been established from an informational argument proposed by Pavlowski *et al.* (use of the "Information Causality Principle") and reformulated by Bub [7-8]. Could this negative result be confirmed more directly, independently of the latter informational considerations, which are not unanimous [9-10]? I think it is the case. The rest of this article will confirm this negative response *from purely descriptive or "linguistic" considerations*, by referring to the properties required by any reliable description of bipartite no-signaling correlations involving incompatible observables.

## II. Which language for describing PR-boxes?

PR-boxes, which, like quantum phenomena, involve stochastic processes and give rise to the loss of information relative to previous outcomes also involve *both incompatible and compatible observables.* The incompatible actions "pushing the joystick at Right" and “pushing the joystick at Left” (incompatible since they cannot be performed at the same time) define *incompatible observables* $A_R$ and $A_L$ (with possible outcomes 0 and 1) that cannot be measured conjointly. Correspondingly, the conjunction or the disjunction of the associated experimental sentences has no experimental meaning. In contrast, if *any* exchange of information between Alice and Bob is ruled out, in agreement with (NS) condition, the game under consideration assumes that the operations performed by Alice *do not perturb* the operations performed by Bob *and vice versa*. This means that these operations are independent and that the couple of observables $A_x$ and $B_y$ (for all possible actions x and y) *can always be evaluated conjointly*. In this case, the conjunction of the two correspondent descriptive sentences can be built and has an experimental counterpart.

The difference between the incompatibility of each of the player’s alternative actions and the compatibility of all the actions of one player with all the actions of the other *must be expressed in an accurate description of this game*. However, this difference cannot be made if the quantities $A_x$ and $B_y$ from which are computed the correlators in equation (1) are regarded as mere random observables (that is, as functions). For, in this case the combination of local operations (in each of the subsystems) and the combination of joint operations (involving both subsystems) are represented without any difference (by a commutative product of functions). Yet, this difference is clearly expressed in terms of structure: due the previously mentioned incompatibility of the two possible alternative operations for each of the players, the structure of the experimental propositions that describe their possible operations, *is not Boolean algebras* but it is a *partial*[1] *Boolean algebra*. Only the sub-algebra corresponding to sentences relative to compatible operations are Boolean[2].

## III. From the partial Boolean algebra of descriptive sentences to the non-commutative algebra of operators.

As is well known since Birhkoff’s and von Neumann’s work on quantum logic and their more recent developments [11-13], such a partial Boolean algebra of experimental sentences[3] is isomorphic to the set of closed subspaces $C(\mathcal{H})$ of an appropriate Hilbert space $\mathcal{H}$ partially ordered by inclusion and endowed with intersection, direct sum and orthogonal complementation:

[1] The term « partial » refers here to the fact that some couples of sentences cannot be combined due to their incompatibility –and not to the fact that the order relation is partial.

[2] Like in quantum logic, the distributivity law for the descriptive sentences of this game is not satisfied, due to the incompatibility of some couples of observables.

[3] Or, equivalently, the orhtocomplemented orthomodular lattice they form.

$$E = [C(\mathcal{H}), \subseteq, \cap, \oplus, \perp],$$

which is, in turn, isomorphic to the corresponding *non-commutative algebra of projectors* onto these subspaces:

$$\mathbb{P} = [\mathrm{Proj}, \leq, \wedge, \vee, {}^{\perp}],$$

where the order relation "$\leq$" is defined by: $P1 \leq P2$ when $P2.P1 = P1$, the operation "$\wedge$" is the usual multiplication of operators (projectors here), "$\vee$" being defined, for commutative projectors (that is within Boolean sub-algebra), as $P1 \vee P2 = P1 + P2 - P1\, P2$, and $P^{\perp}$ being the projector onto the orthogonal complementation of Im(P).

Accordingly, unlike the Boolean structure of classical experimental sentences of classical mechanics to which Kolmogorovian probabilities can be assigned, the latter non-commutative algebra of projectors *imposes* a different, non-classical probability measure. This probability measure, *which is the only possible*, is provided by Gleason's theorem [14] -or, equivalently, by Born's rule- which can be written, for a given vector state **v** of $\mathcal{H}$: $p(a_i/x) = \mathbf{v}.\, P_i\, \mathbf{v}$, where $P_i$ is the projector onto the subspace associated with the outcome $a_i$ for the measurement "x".

**IV. Super-quantum, no-signaling correlations cannot then exist.**

From the previous results, we can assert that bipartite no-signaling correlations that involve incompatible local observables cannot be represented within a commutative algebra of functions but *requires* to be represented in a non-commutative algebra of projectors on an appropriate vector space. In contrast to what is generally thought, we have *no choice*: representing this experimental situation within such a non-commutative algebra of operators is *a logical necessity* due to the structure of its descriptive sentences. Consequently, the correlators of equation (1) cannot be computed by representing observables by mere random variables but by *operators* of a C*-algebra. In the aforementioned case of bipartite correlations, the observable $A_{xi}$, with $x_i = 0$ or 1, referring to the choice of the input $x_i$ by Alice is a linear Hermitian operator acting on the two-dimensional Hilbert space spanned by its eigenvectors $|0>_{Axi}$ and $|1>_{Axi}$, the corresponding projectors being $|0><0|_{Axi}$ and $|1><1|_{Axi}$ –and similarly for the observable $B_{yj}$.

Now, as is well known, within such a non-commutative algebra of observable-operators, no-signaling correlations between binary observables with outcomes $\pm$ 1 are constrained by the Tsirelson bound [15-16]. No-signaling super-quantum correlations cannot then exit for structural reasons; no mysterious "superselection" principle is required for explaining their inexistence.

Super-quantum correlations could only be observed if the no-signaling condition (NS) is removed, that is, either in the trivial case the two subsystems are not separated (they interact) *or if non-local (or non-spatial) communication can take place between them*, that is, *if they can exchange information non-locally*. Consequently, if such super-quantum correlations for spatially separated subsystems on which are defined incompatible observables are observed, this would then show that non-local communication is possible and challenge the validity of quantum theory, since the latter shows its impossibility (no-communication theorem).